\documentclass[aps,reprint, superscriptaddress, nofootinbib]{revtex4-1}

\usepackage{amsmath, amssymb, amsthm}
\usepackage{bm,bbm}

\usepackage{graphicx}
\usepackage{hyperref}
\usepackage[margin=2cm]{geometry}
\usepackage{xcolor}
\usepackage{qcircuit}
\usepackage{multirow}
\theoremstyle{plain}

\newcommand{\be}{\begin{equation}}
\newcommand{\ee}{\end{equation}}
\newcommand{\bpm}{\begin{pmatrix}}
\newcommand{\epm}{\end{pmatrix}}

\newcommand{\ket}[1]{\ensuremath{\left| #1 \right \rangle}}
\newcommand{\bra}[1]{\ensuremath{\left \langle #1 \right |}}
\newcommand{\braket}[2]{\ensuremath{\left\langle #1\left|#2 \right.\right\rangle}}
\newcommand{\ketbra}[2]{\ket{#1}\bra{#2}}

\newcommand{\tr}{\mathrm{tr}}

\newtheorem{defi}{Definition}

\begin{document}
\title{Quantum embeddings for machine learning}
\author{Seth Lloyd}
\affiliation{Massachusetts Institute of Technology, 77 Massachusetts Avenue, Cambridge, MA 02139, USA}
\affiliation{Xanadu, Toronto, Canada}
\author{Maria Schuld}
\affiliation{Xanadu, Toronto, Canada}
\author{Aroosa Ijaz}
\affiliation{Xanadu, Toronto, Canada}
\author{Josh Izaac}
\affiliation{Xanadu, Toronto, Canada}
\author{Nathan Killoran}
\affiliation{Xanadu, Toronto, Canada}

\date{\today}
\begin{abstract}
Quantum classifiers are trainable quantum circuits used as machine learning models. The first part of the circuit implements a quantum feature map that encodes classical inputs into quantum states, embedding the data in a high-dimensional Hilbert space; the second part of the circuit executes a quantum measurement interpreted as the output of the model. Usually, the measurement is trained to distinguish quantum-embedded data. We propose to instead train the first part of the circuit---the embedding---with the objective of maximally separating data classes in Hilbert space, a strategy we call \textit{quantum metric learning}. As a result, the measurement minimizing a linear classification loss is already known and depends on the metric used: for embeddings separating data using the $\ell_1$ or trace distance, this is the Helstr\o{}m measurement, while for the $\ell_2$ or Hilbert-Schmidt distance, it is a simple overlap measurement. 
This approach provides a powerful analytic framework for quantum machine learning and eliminates a major component in current models, freeing up more precious resources to best leverage the capabilities of near-term quantum information processors. 
\end{abstract}

\maketitle

Machine learning is a potential application for near-term intermediate scale quantum computers \cite{wittek2014quantum, biamonte2017quantum, schuld2018supervised}. Quantum machine learning algorithms have been shown to provide speed-ups over their classical counterparts for a variety of tasks, including principal component analysis, topological data analysis, and support vector machines \cite{lloyd2014quantum, lloyd2016quantum, rebentrost2014quantum}.
This paper investigates quantum machine learning using variational quantum classifiers, which are parametrized quantum circuits that embed input data in Hilbert space and perform quantum measurements to discriminate between classes \cite{farhi2018classification, schuld2018circuit, benedetti2019parameterized}. The variational strategy can be extended to more complex classification tasks, hybrid quantum-classical models, and generative models \cite{mari2019transfer, benedetti2019generative}. The training of the circuit is performed by optimizing the parameters of quantum gates -- such as the angles of Pauli-rotations -- with a hybrid quantum-classical optimization procedure \cite{mitarai2018quantum,schuld2019evaluating}. It has recently been shown \cite{schuld2019quantum, havlivcek2019supervised} that this strategy represents a quantization of classical \textit{kernel methods} such as support vector machines \cite{scholkopf2001learning}, which implicitly embed data in a high-dimensional Hilbert space and find decision hyperplanes that separate the data according to their classes.   

In \cite{schuld2019quantum, havlivcek2019supervised}, the quantum feature map that embeds the data is taken to be a fixed circuit, and the adaptive training is performed on a variational circuit that adapts the measurement basis.  By contrast, here we note that if the data is ``well-separated'' in Hilbert space, the best measurements to distinguish between classes of data are known and can be performed with shallow quantum circuits. The metric under which the data is separated defines which measurement should be performed: the best measurement for data separated by the trace distance ($\ell_1$) is the Helstr\o{}m minimum error measurement \cite{helstrom1976quantum}, and the best measurement for the Hilbert-Schmidt ($\ell_2$) distance is obtained by measuring the fidelity or overlaps between embedded data, for example with a simple swap test. We present efficient circuits to implement both measurements, which in the case of the fidelity measurement is extremely short. 

\begin{figure}[t]
\includegraphics[width=0.48\textwidth]{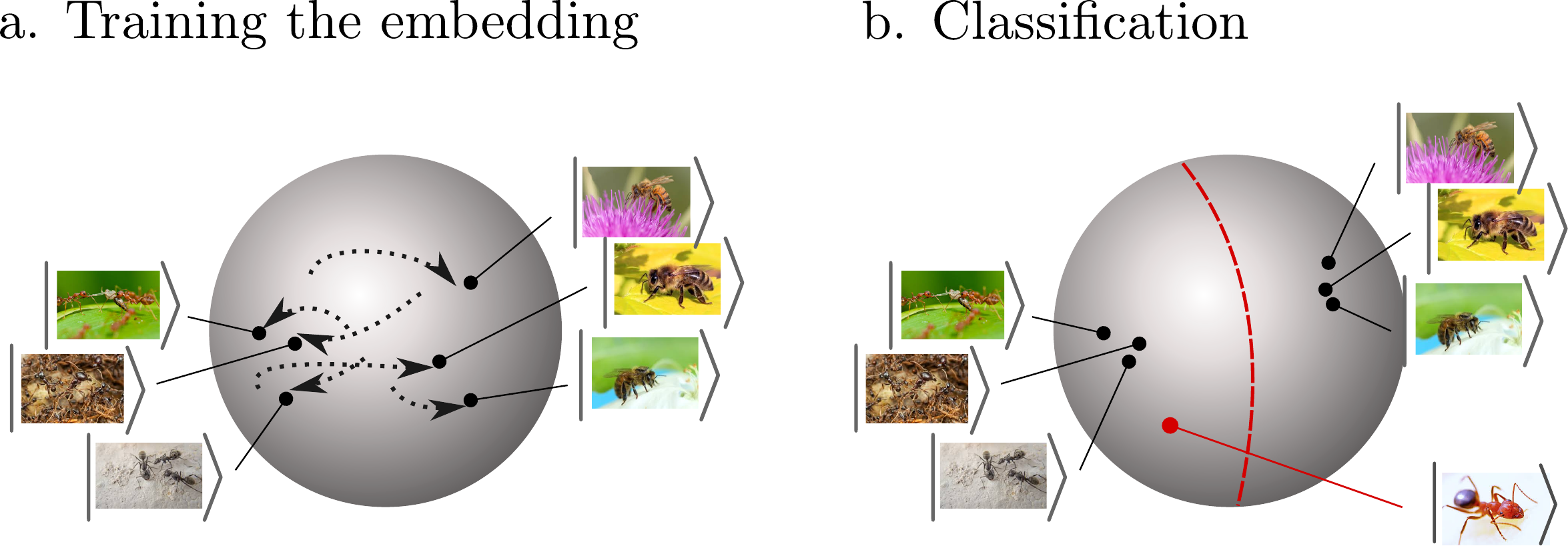}
\caption{Illustration of quantum metric learning. a. The embedding is trained to maximize the distance of the data clusters in the Hilbert space of quantum states. b. The measurement used to classify new inputs depends on the distance measure used. The simple decision boundary in Hilbert space can correspond to a highly complex decision boundary in the original data space.}
\label{Fig:fig0}
\end{figure}

Knowing the optimal quantum measurements for discriminating between clusters of embedded data has important consequences; the common approach of training a variational circuit after the embedding and before the measurement -- and thereby the bulk of the computational resources spent on the quantum classifier -- becomes obsolete. We argue that instead, the adaptive training of the quantum circuit should be focused on training a quantum feature map that carries out a maximally separating embedding (see Figure \ref{Fig:fig0}). This approach is known as ``metric learning'' in the classical machine learning literature \cite{bromley1994signature, chopra2005learning}, where feature maps, and thereby a metric on the original data space, are learned with models such as deep neural networks. Note that while deep metric learning extracts low-dimensional representations of the data, quantum computing allows us to learn high-dimensional representations without explicitly invoking a kernel function. 

We numerically investigate adaptive methods for training quantum embeddings using the PennyLane software package for hybrid optimization \cite{bergholm2018pennylane}, and explore the performance of the  measurements for small examples. We note that due to potential quantum advantages \cite{harrow2017quantum}, the embeddings performed by quantum circuits can be inaccessible to classical computers, and close with a brief discussion of the experimental feasibility of implementing quantum embeddings on existing and near-term quantum computers.

\section{Quantum embeddings}

Kernel methods for learning operate by embedding data points $x$
as vectors $\vec x \in {\it H}$ into a Hilbert space $H$,
a vector space with an inner product.   The data is
then analyzed by performing linear algebra on the embedded  vectors.  Standard results on metric spaces imply that
any finite metric space can be embedded in a sufficiently
high-dimensional Hilbert space in such a way that the
metric on the Hilbert space faithfully approximates the metric of the original space (e.g., \cite{bourgain1985lipschitz}). The goal of the embedding process is to find a representation of the data such that the known metric of the Hilbert space faithfully reproduces the unknown metric of the original data, for example, the human-perceived ``distance'' between pictures of ants and pictures of bees. For such a faithful embedding to be possible, the dimension of the Hilbert space may have to be large -- at least the order of the number of data points. If one can find a faithful embedding, the computations required to compare data vectors and assign them to clusters can be performed using standard linear algebraic techniques.

Quantum computers are uniquely positioned to perform kernel methods. The states of quantum systems are vectors in a high-dimensional Hilbert space, and standard linear algebraic operations on that Hilbert space can often be performed in time poly-logarithmic in the dimension of the space -- an exponential speed up over the corresponding classical operations. The Hilbert space of an $n$-qubit quantum computer is the vector space ${\cal C}^N$, where $N=2^n$, and the states $\psi, \varphi$ are written in Dirac notation as $| \psi \rangle, | \varphi \rangle$ with the inner product $ \psi^\dagger \varphi = \langle \psi|\varphi\rangle$. Because of the probabilistic interpretation of quantum mechanics,
state vectors are normalized to 1. 
A measurement to verify that a vector $\ket{\varphi}$ is equal to $|\psi\rangle$ corresponds to a projection $ P_{\psi} = |\psi\rangle\langle \psi|$, which yields the answer Yes with probability $\langle \phi | P_{\psi} | \phi\rangle = | \langle \psi | \phi\rangle|^2$. 
The process of uniformly sampling $M$ quantum states from a set $\{\ket{\psi_i}\}$ is described by the density matrix $\rho = \frac{1}{M}\sum_i \ketbra{\psi_i}{\psi_i}$. The $\ell_1$ distance between the statistical ensemble $\rho$ and another ensemble $\sigma = \frac{1}{M'}\sum_j \ketbra{\phi_j}{\phi_j}$ is the widely studied trace distance $D_{\mathrm{tr}}(\rho, \sigma) = \frac{1}{2}\tr( |\rho - \sigma| )$, while the $\ell_2$ distance is known as the Hilbert-Schmidt distance $D_{\mathrm{hs}}(\rho, \sigma) = \tr( (\rho - \sigma)^2 )$.

Quantum computers also naturally map data into Hilbert space. The map that performs the embedding has been termed a \textit{quantum feature map} \cite{schuld2019quantum, havlivcek2019supervised}. A quantum embedding is then the representation of classical data points $x$ as quantum states $\ket{x}$, facilitated by the feature map. We assume here that the quantum feature map is enacted by a quantum circuit $\Phi(x, \theta)$ which associates physical parameters in the preparation of the state $\ket{x} = \Phi(x, \theta) \ket{0 \dots 0}$ with the input $x$. Another set $\theta$ of physical parameters is used as free variables that can be adapted via optimization. All parameters of the embedding are classical objects -- they are simply numbers in a suitable high-dimensional vector space — and can be trained using classical stochastic gradient descent methods \cite{kubler2019adaptive, sweke2019stochastic}. For example, in a superconducting or ion-trap quantum computer, training could specify the time-dependent microwave or laser pulses which are used to address the interacting qubits in the device; in an optical quantum computer the parameters could be squeezing amplitudes or beam splitter angles. 
 
In the following, we will limit ourselves to binary classification for simplicity\footnote{Note that any binary classifier can be turned into a multi-class classifier, for example using a one-versus-all strategy.}. In this task, we are given a dataset with examples from two classes $A$ and $B$ of labels $1$ and $-1$, respectively, and have to predict the label of a new input $x$, which may or may not be from the training data set of examples. Uniformly sampling $M_a$ inputs from class $A$ and embedding them into Hilbert space prepares the ensemble $\rho = \frac{1}{M_a}\sum_{a \in A} |a\rangle \langle a|$, while sampling $M_b$ inputs from class $B$ prepares $\sigma = \frac{1}{M_b}\sum_{b \in B} |b\rangle \langle b|$.

\section{Optimal measurements}

A quantum classifier predicts a label for an input $x$ mapped to $\ket{x}$ via a quantum measurement. In general, the result of the classifier can be written as $f(x) = \bra{x} \mathcal{M} \ket{x} = {\rm tr} (\ket{x} \bra{x} \mathcal{M} )$, where $\mathcal{M}$ is the quantum observable corresponding to the measurement, represented by a hermitian operator (see Appendix \ref{App:defs}). The continuous expectation can be turned into a binary label by thresholding on a value $\tau$. The decision boundary of a quantum classifier on the space
of density matrices is the hyperplane defined
by ${\rm tr} (\ket{x} \bra{x} \mathcal{M}) = \tau$. In the following we assume $\tau = 0$. This separating
hyperplane is the quantum analogue of a classical support vector machine.

According to statistical learning theory \cite{vapnik1999overview}, the quality of a classifier $f$ is measured by the expected risk. Given a training set  $\{ (x^m, y^m)\}_{m=1}^M$ of input-output data pairs from classes $A$ and $B$, the expected risk is typically approximated by the empirical risk $\hat{I}[f] =  \frac{1}{M}\sum_{m} L(f(x^m), y^m)$, where $L$ is a loss function indicating how similar the output $f(x^m)$ is to the target label $y^m$. Choosing the linear loss function\footnote{The more common quadratic loss has additional terms $f(x)^2$ which are of order $1/M^2$ and punish large values of $f$. For typical quantum classifiers $f(x)$ lies in a bounded interval, and the linear loss is a more natural choice. Other loss functions often depend on the linear loss.} $L(f(x), y) = -f(x)y$, the empirical risk of a quantum classifier is given by $\hat{I}[f] = -\tr ( (\rho- \sigma)  \mathcal{M})$, where $\rho$ is the ensemble of inputs from class $A$ with labels $1$, and $\sigma$ is the ensemble of inputs from class $B$ with label $-1$. 

We consider two well-known measurements from quantum information theory, a fidelity as well as a Helstr\o{}m measurement. The fidelity measurement can be implemented by a SWAP test, which is a simple quantum procedure that estimates the overlap or fidelity $ F =| \langle \varphi| \psi\rangle|^2$ of two quantum states $|\varphi\rangle$, $|\psi\rangle$. Performing the SWAP test on $|\varphi\rangle$, $|\psi\rangle$ for $k$ times allows one to estimate the fidelity to an accuracy $\pm \sqrt{ F(1-F)/k}$. The SWAP test on $n$-qubit states can be performed using a quantum circuit that adjoins a single ancilla qubit to the $2n$ qubits used to represent $|\varphi\rangle$ and $|\psi\rangle$, and by performing $n$ controlled SWAP operations, followed by measurement of the ancilla qubit. For example, if we embed our data in a $2^{50} \approx 10^{15}$ dimensional Hilbert space, each SWAP test requires $50$ controlled SWAP operations to be performed on $101$ qubits. As shown in Ref. \cite{havlivcek2019supervised}, if the embedding circuit $\Phi(x)$ can be inverted, one can alternatively implement a circuit $\Phi(x')^{\dagger} \Phi(x) \ket{0 \dots 0}$ on a register of $n$ qubits and for two inputs $x, x'$, and measure the overlap to the $\ket{0 \dots 0}$ state. While requiring the same amount of repetitions, this ``inversion test'' scheme only needs $n$ qubits overall.

A \textit{fidelity classifier} (see Appendix \ref{App:fid}, but also \cite{blank2019quantum, schuld2017implementing}) uses either the SWAP or inversion test to assign a state $|x\rangle$ to $A$ if the average probability of projecting the state onto the training states in $A$ is higher than the probability of projecting it onto the training states in $B$. Hence, $f_{\mathrm{fid}}(x) = \langle x | (\rho - \sigma) |x\rangle$. The empirical risk of the fidelity classifier (using linear loss)  is given by $\hat{I}[f_{\mathrm{fid}}] = -\tr ( (\rho- \sigma)^2) = - D_{\mathrm{hs}}(\rho, \sigma)$. \textit{Therefore, maximizing the Hilbert-Schmidt distance of the embedding minimizes the empirical risk of the fidelity classifier under linear loss.}

The second measurement we consider is a Helstr\o{}m measurement. Helstr\o{}m measurements are known to be optimal for the task of quantum state discrimination, in which a single measurement has to assign a quantum state to one of two clusters \cite{helstrom1976quantum, chefles2000quantum}, a fact that has been used in the context of machine learning before \cite{gambs2008quantum, sergioli2017quantum, sentis9unsupervised}. The classification problem we consider here differs from this setting, since the classifying measurement can be repeated, and the objective is to minimize expected risk (see Appendix \ref{App:state_dis}). A Helstr\o{}m measurement projects onto the positive and negative subspaces of $\rho-\sigma$, using projection operators $\Pi_+$, $\Pi_-$, so that $f_{\rm hel}(x) = \bra{x}\Pi_+ - \Pi_- \ket{x}$. 

As shown in \cite{lloyd2014quantum}, the minimum error probability measurement can be performed efficiently (in time $O(R n)$) on a $n$-qubit quantum computer as long as $\rho$ and $\sigma$ have a rank $R$ that does not grow with the dimension of the Hilbert space. In Appendix \ref{App:hel} we outline a quantum circuit implementation and show that in a single run of the measurement circuit, $\rho$ and $\sigma$ are pure states, so that $R \leq 2$. The empirical risk of the corresponding \textit{Helstr\o{}m classifier} under linear loss is given by the trace distance, $\hat{I}[f_{\mathrm{hel}}] = -\tr ( |\rho- \sigma|) = -2D_{\mathrm{tr}}(\rho, \sigma)$, and hence \textit{maximizing the trace distance of the embedding minimizes the risk of the Helstr\o{}m classifier under linear loss}. 

\begin{figure}[t]
\includegraphics[width=0.48\textwidth]{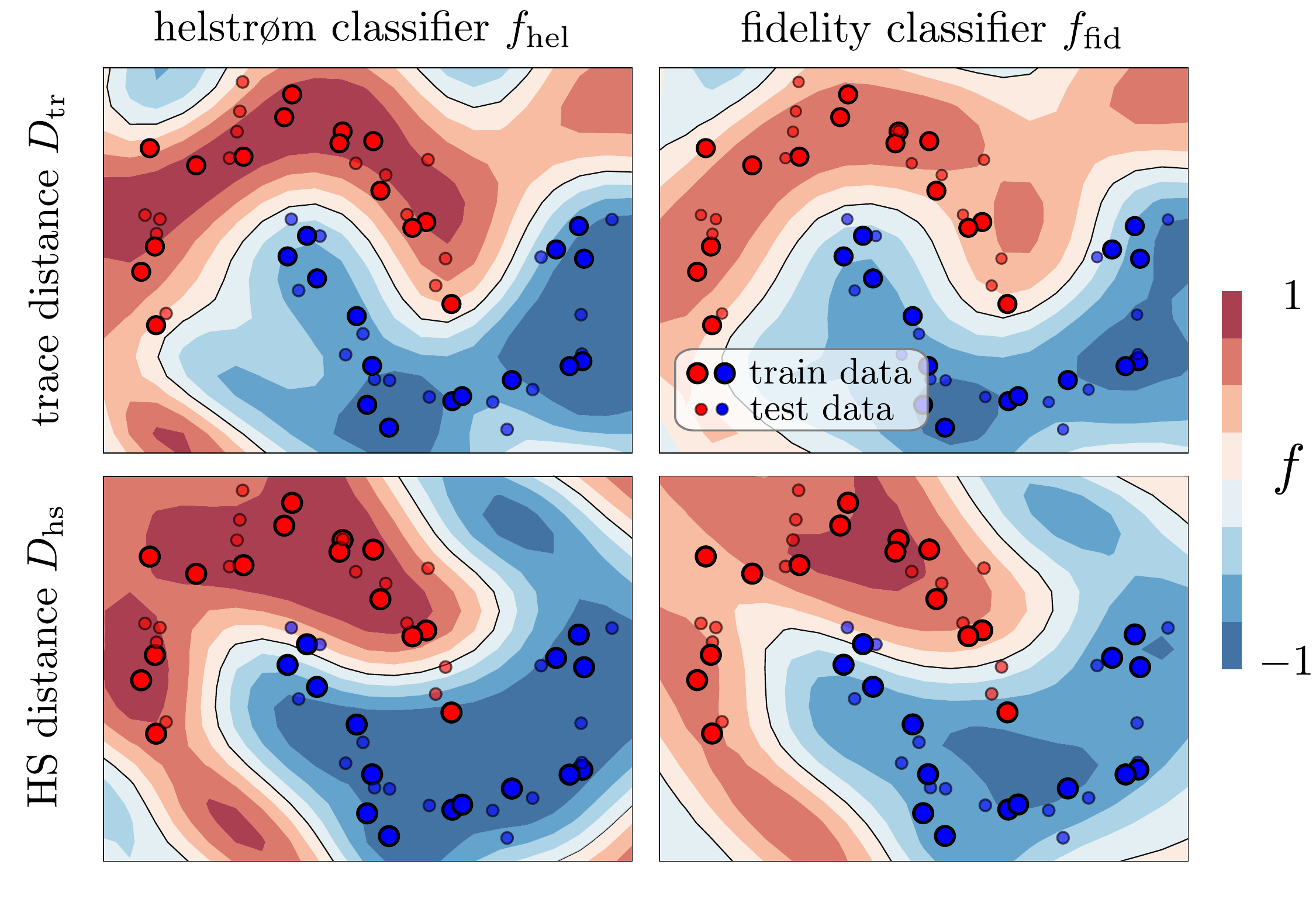}
\caption{Decision boundary on a $2$-d moons dataset for the Helstr\o{}m and fidelity classifiers, training the embedding with the trace and Hilbert-Schmidt distance, respectively. The embedding was trained for $500$ steps with an RMSProp optimizer and batch size $5$, using a $2$ qubit QAOA feature map of $4$ layers, and reaching a final cost of $0.28$ for $\ell_1$ and $0.55$ for $\ell_2$ training (see Section \ref{Sec:practical}). Although the details of the plots vary with the hyperparameters of the training, the example illustrates that both classifiers give rise to an overall similar, but not identical decision boundary. Consistent with state discrimination theory, one can also see that the Helstr\o{}m classifier has outputs closer to the extremes $1$ and $-1$. The embedding was trained in PennyLane, and the classifiers were simulated analytically.}
\label{Fig:fig1}
\end{figure}

In summary, once one has decided which Hilbert space metric to use when training the embedding, the optimal measurement for state assignment with respect to the linear loss is known. Figure \ref{Fig:fig1} shows the decision boundaries of the two classifiers for a $2$-d moons dataset, for which quantum embeddings were separately trained with the $\ell_1$ and $\ell_2$ distance. 

\section{ Training the embedding in practice}

While the trace distance plays a central role in quantum information theory, the Hilbert-Schmidt distance can be measured and hence optimized by a much smaller quantum circuit, which is significant for near-term quantum computing. Accordingly, in our analysis we focus on the Hilbert-Schmidt distance only. Note that the distances are closely related by \cite{coles2019strong} $\frac{1}{2}D_{\mathrm{hs}} \leq D_{\mathrm{tr}}^2 \leq r D_{\mathrm{hs}}$, where $r = \mathrm{rank} (\rho) \mathrm{rank} (\sigma) / (\mathrm{rank} (\rho) + \mathrm{rank} (\sigma))$. If $\mathrm{rank} (\rho) = \mathrm{rank}( \sigma) = 1$, we have the equality $D_{\mathrm{tr}}^2 = \frac{1}{2} D_{\mathrm{hs}}$.  

In Hilbert-Schmidt optimization, the cost to be minimized during training is given by $C=1- \frac{1}{2}D_{\rm hs}$. The different terms $ \tr \rho \sigma$, $\tr \sigma^2$ and $\tr \rho^2$ in $D_{\rm hs}$ can be estimated on a quantum computer by performing repeated SWAP or inversion tests, either between inputs of the same class, or between inputs of different classes. 
The terms in $C$ have a simple intuitive meaning. The term $\tr \rho \sigma$ measures the distance between the two ensembles in Hilbert space via the inter-cluster overlap;  $\tr \rho \sigma = 1$ indicates that the ensembles are constructed from the same set of pure states, while $\tr \rho \sigma = 0$ means that all embedded data points are orthogonal. 
The `purity' terms ${\rm tr} \rho^2, {\rm tr} \sigma^2$ are measures for the intra-cluster overlap, which is closely related to the rank of the respective density matrices. When ${\rm tr} \rho^2 = 1$, the embedding maps all inputs $\{a\}$ to the same pure state $\ket{a}$, and hence ${\rm rank}(\rho) = 1$; this means that the cluster of class-$A$ states in Hilbert space is maximally tight. For ${\rm tr} \rho^2 = 1/2^n$ ($n$ being the number of qubits representing $\ket{x}$) the density matrix has full rank and the states ${\ket{a}}$ are maximally spread in Hilbert space. Overall, the cost lies in the interval $[0, 2]$. 

Optimizing according to Hilbert-Schmidt distance has some useful features: it automatically leads to low-rank embeddings which correspond to tight clusters in Hilbert space.  In addition to reducing the difference between $\ell_1$ (Helstr\o{}m) and $\ell_2$ (Hilbert-Schmidt) optimization, low-rank $\rho, \sigma$ are favourable for generalization, since they increase the similarity of an unseen data point to its cluster. Low-rank data clusters also reduce the number of samples $S$ required to execute the measurement: the probability of assigning a class $A$ input to class $A$ is given by $p = \tr \rho^2$ while assigning it to class $B$ is $q = \tr \rho \sigma$. To distinguish between the probabilities $p, q$ we require $\sqrt{(p (1-p))/S} < p-q$ as an upper bound for the error, so that  $S > \mathcal{O}(\frac{1}{\tr{\rho^2}})$.

 Lastly, the Hilbert-Schmidt distance is equivalent to the \textit{maximum mean discrepancy} between the two distributions from which the data from the two classes was sampled \cite{gretton2012kernel}. This measure has proven a powerful tool in training Generative Adversarial Networks \cite{li2015generative, li2017mmd}, and opens a wealth of statistical results to training \cite{tolstikhin2016minimax}.

\section{ Numerical experiments}\label{Sec:practical}

\begin{figure}[t]
\includegraphics[width=0.45\textwidth]{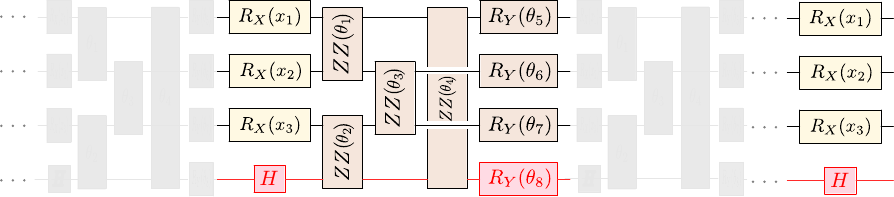}
\caption{Ansatz for a single trainable layer of the embedding used in the experiments, showing $3$ inputs $x_1, x_2, x_3$, as well as trainable ZZ-entanglers and $R_Y$ rotations. Latent qubits, as shown with the fourth qubit (red), can be added to increase the dimension of the Hilbert space. The quantum embedding consists of several such layers, and a final repetition of the feature encoding rotations.}
\label{Fig:fig2}
\end{figure}

We demonstrate the ideas of trainable quantum embeddings with the PennyLane software framework \cite{bergholm2018pennylane}, and an embedding circuit ansatz that is inspired by the Quantum Approximate Optimization Algorithm (QAOA) \cite{farhi2014quantum}\footnote{This has also -- and more fittingly -- been termed the ``Quantum Alternating Operator Ansatz''.}. Using repetitions or ``layers'' of the ansatz (see also \cite{perez2019data}) can implement classically intractable \cite{lloyd2018quantum} feature maps that are universal for quantum computing. The ansatz uses one- and two-qubit quantum logic operations, such as single qubit rotations $e^{-i\theta_1 \sigma_{\hat j}}$ and two qubit $ZZ$ interactions $e^{-i \theta_2 \sigma_z\otimes \sigma_z}$, where the axes of rotation $\hat j$ and the angles of rotation $\theta_{1,2}$ are individual parameters that we can select. Here we will consider a circuit consisting of layers of individually programmable single qubit Pauli-X rotations $R_x$, and a chain of individually programmable pairwise $ZZ$ or ``Ising'' interactions (which commute with each other). We have empirically found it favourable to also include a block of single qubit Pauli-Y rotations $R_y$ as ``local fields''.\footnote{Note that fully-connected nearest-neighbor coupled pairwise $ZZ$ interactions in one or more dimensions suffice for universality \cite{lloyd2018quantum}.} The feature-encoding gates are repeated once more after the last layer. The input to the circuit is taken to be the state $|0\ldots 0\rangle$, and one can increase the feature-space dimension by adding ``latent qubits'' as shown in Figure \ref{Fig:fig2}.

\begin{figure*}[t]
\includegraphics[width=\textwidth]{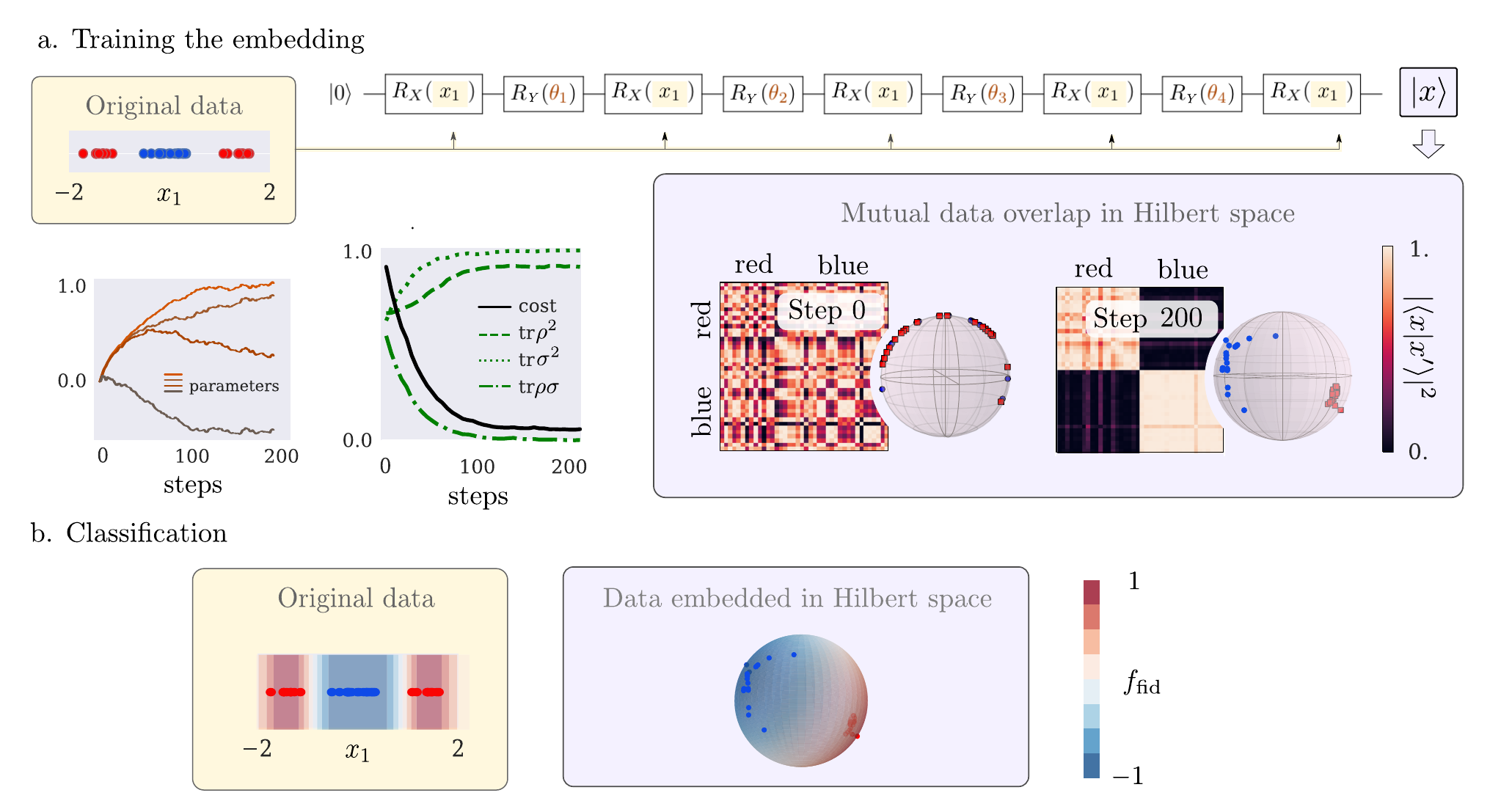}
\caption{Illustrative example of a quantum embedding and classification for a non-overlapping, but not linearly separable, one-dimensional dataset. Training was done using the cost $C$ defined in Section \ref{Sec:practical}, and an RMSProp optimizer with initial learning rate $0.01$ and batch size $2$. a. The untrained feature map distributes the data arbitrarily on the Bloch sphere, while after $200$ steps of training the classes are well-separated. b. The fidelity classifier draws a linear decision boundary on the bloch sphere, which translates to two linear decision boundaries in the original space. The simulations were done in the PennyLane software framework \cite{bergholm2018pennylane}.}
\label{Fig:fig3}
\end{figure*}

To perform the embedding, we designate the $R_x$ parameters to encode the input features $x = (x_1,...,x_N)^T$, and the remainder to encode the trainable parameters $\theta$. The overall unitary transformation $\Phi(x, \theta)$ is then a function of the weights and the input, and the embedding takes the form $ x\rightarrow |x\rangle = \Phi(x, \theta) |0\ldots 0\rangle$. The overlap between two embedded states is $\langle x_1 | x_2 \rangle$.
Because of universality of the circuit ansatz in the limit of many layers, evaluating this overlap for different values of $\theta,x_1, x_2$ is equivalent to evaluating the outcome of an arbitrary quantum computation over $n$ qubits, and so is inaccessible to a classical computer unless BQP=P. Even very shallow circuits taken from this circuit model suffice to produce embeddings that cannot be produced classically. A three-layer circuit consisting of $X$ rotations, followed by an  Ising layer and a second layer of $X$ rotations furthermore implements instantaneous quantum polynomial-time computation (IQP) \cite{bremner2010classical}. The fact that quantum systems can explore a larger class of embeddings than classical systems gives a motivation to explore whether this larger class of embeddings might allow quantum computers to perform better on some set of classification problems.

\begin{figure*}[t]
\includegraphics[width=\textwidth]{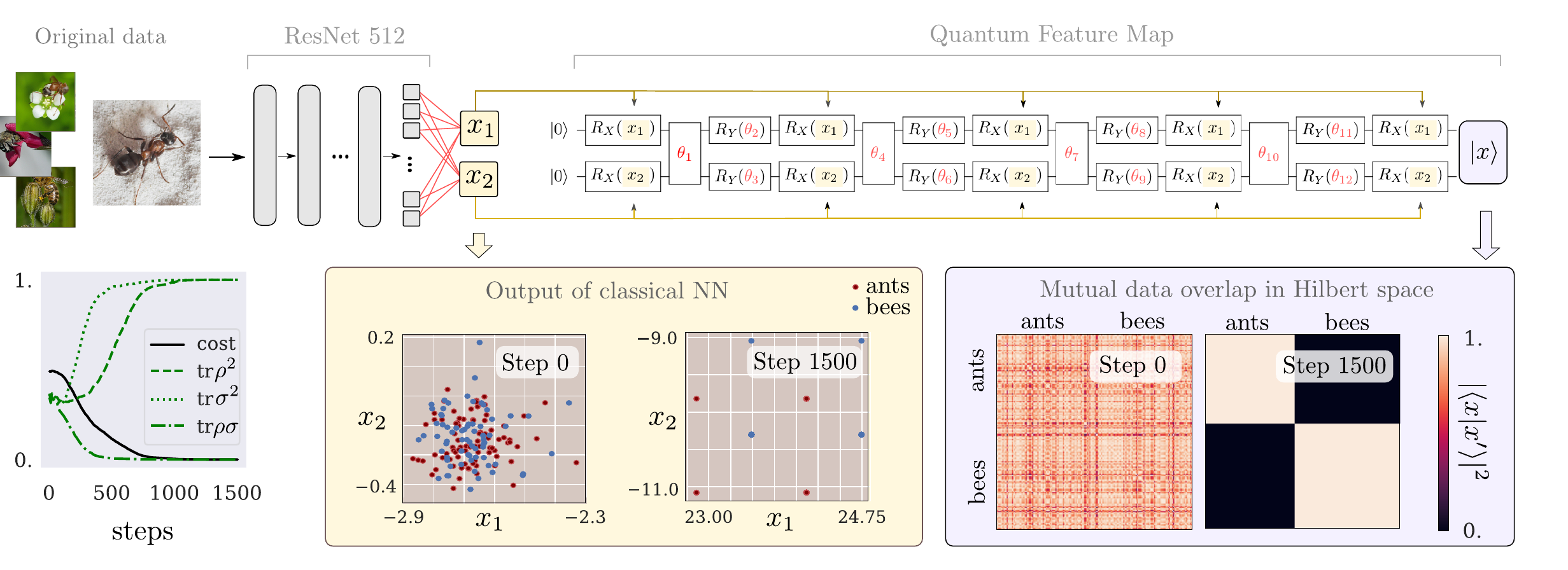}
\caption{Hybrid quantum-classical embedding. An image is fed into a pre-trained ResNet. The last layer is replaced by a linear layer, transforming the $512$-dimensional output to a $2$ dimensional feature vector, which is fed into a parametrized quantum circuit. The circuit parameters are trained together with the final classical layer (red). After $1500$ training steps with an Adagrad optimizer, the classical weights learn to map the two classes to periodically arranged points in the intermediate feature space, which allows the quantum circuit to perfectly separate the two classes in quantum feature space. Each optimization step uses a batch of $2$ randomly sampled training points. The simulations were done in PennyLane.}
\label{Fig:fig4}
\end{figure*}

Figure \ref{Fig:fig3} shows a toy model of embedding $1$-dimensional data into a single qubit using the ``QAOA feature map''. The trained feature map embeds the data into tight and linearly separable clusters on the Bloch sphere, and a linear decision boundary in the $2$-dimensional Hilbert space corresponds to two linear decision boundaries in the original space. Figure \ref{Fig:fig4} shows that the quantum embedding can be used on larger datasets by a hybrid approach, adapting an experiment from \cite{mari2019transfer} that combines a classical ResNet with a quantum circuit. PennyLane allows the joint optimization of parameters for the classical and quantum model. Interestingly, one sees that the neural network learns to map the data to a periodic structure, which is a very natural input to a quantum circuit with $R_x$ feature encoding. 

While these demonstrations show that the training of quantum embeddings is possible, in this study we do not attempt to answer the question of whether quantum embeddings are better than classical machine learning methods, which is in some sense as hard a question as the general power of quantum machine learning. However, our approach offers a fruitful analytical framework: the power of a quantum classifier translates into the ability to find embeddings that maximize the distance between data clusters in Hilbert space.

\section{ Capacity of embeddings on near-term quantum devices}

Lastly, we briefly discuss the feasibility of implementing quantum embeddings
on near-term quantum computers such as superconducting or ion-trap quantum
information processors.   We require $2n + 1$ qubits to embed two
$n$-qubit states and to perform a swap test.   Each $n$-qubit state is
created by applying a pulse sequence to our qubits, specified by
parameters $x, \theta$, as above, where $x$ specifies the input data
and $\theta$ determines the form of the embedding. Applying this pulse
sequence to an initial state $|0\ldots 0\rangle$ yields an
embedded state $|x\rangle = \Phi(x,\theta) |0\ldots 0\rangle$. 
Let us consider the physical requirements for embedding a single
$n$-qubit state.   

Assume that we can address each of the $n$ qubits
in our device individually by time-varying electromagnetic pulses with 
bandwidth $\Omega$.  The number of classical bits of information that the pulses can embed in the quantum state of the computer
within the coherence time $t$ is then $2\Omega t n b$ where $b$ is the number of bits required to specify the strength of the electromagnetic
field at each sample time. To satisfy the requirements of quantum
intractability, we ask that the circuit be of depth $O(n)$, to spread
quantum information throughout the device via qubit-qubit interactions.

To put in some current numbers, for $100$ superconducting qubits with $\Omega$ equal to $10$ GHz, $10$ bits required to specify the strength of the field, and a coherence time of $10^{-3} s$, we can embed $O(10^{10})$ classical bits within the decoherence time. An interaction strength of $100$ MHz suffices to enact a quantum circuit of depth $100$ gates, which is deep enough to allow quantum information to spread through the system. Other proposed solid-state quantum computing platforms, e.g., silicon quantum dots or NV-diamond centers, exhibit similar numbers to superconducting systems.  For $100$ ion-trap qubits with $\Omega = 100$ MHz, $10$ bits per sample, and a coherence time of $1 s$, we can embed $O(10^{11})$ classical bits within the decoherence time. The typical interaction strength of $10$ kHz suffices to enact a quantum circuit with depth $10^4$, which is more than adequate for a quantum embedding that is unattainable classically.
Other proposed atom-optical based quantum computing platforms, e.g., optical
lattices, exhibit numbers similar to ion traps. In conclusion, high-dimensional embeddings of large data sets for quantum
machine learning should be accessible on current devices. 

\section{Conclusion}

This work proposes ``metric quantum learning'' as an experimentally accessible and analytically promising approach to quantum machine learning. Classical data points are embedded as quantum states, and then compared using optimal quantum measurements. The embedding is adjusted adaptively to train the quantum classifiers. Both the embedding and the measurements can be performed on the relatively small, shallow quantum circuits afforded by intermediate-term quantum computing platforms. While the results of quantum embeddings can in general not be reproduced on classical computers, the existence of a useful quantum advantage is an open question.  

\bigskip\noindent{\it Acknowledgments}

The authors want to thank Juan Miguel Arrazola, Joonwoo Bae, Guillaume Dauphinais, Yann LeCun, and Nicolas Quesada for helpful comments.

\onecolumngrid
\setlength\parindent{0pt}
\appendix

\section{Classification with quantum circuits}\label{App:defs}

Here we provide a more technical definition of quantum classifiers in the context of binary classification. Like in the main paper, we consider the following problem setup:

\begin{defi}
Let $X$ be a data domain. Assume we are given data samples $\{ a_1,...,a_{M_a} \}$ from class $A \subseteq X$ with label $1$, and data samples $\{ b_1,...,b_{M_b} \}$ from class $B \subseteq X$ with label $-1$, as well as a new input $x \in X$. The problem of \textbf{binary classification} is to predict the label of $x$, assigning it to either class $A$ or class $B$.
\end{defi}

A classifier is a model or algorithm solving the problem of binary classification:

\begin{defi}
A \textbf{classifier} is a map from the data domain to the real numbers, $f: X \rightarrow \mathbb{R}$. The classifier assigns a binary label to $x$ according to the thresholding rule 
$$ y = \begin{cases} -1 \text{ if } f(x) < \tau \\ 1 \text{ if } f(x) \geq \tau, \end{cases}$$
with $\tau \in \mathbb{R}$. If no other information is provided, $\tau$ is assumed to be zero.
\end{defi}

Quantum classifiers are specific models that use quantum theory to solve classification tasks. They are based on a representation of data as states of a quantum system.

\begin{defi}
A \textbf{quantum embedding} is a quantum state $\ket{x}$ that represents a data input $x \in X$. It is facilitated by a \textit{quantum feature map} $\phi:x \rightarrow \ket{x}$. For example, the quantum feature map can be executed by a quantum circuit $\Phi(x)$ whose gate parameters depend on $x$. 
\end{defi}

A quantum classifier combines an embedding with a measurement.

\begin{defi}
Let $\ket{x}$ be an \textit{embedding} of a data input $x \in X$. A \textbf{quantum classifier} is a classifier where $f(x)$ is an expectation $\bra{x} \mathcal{M}\ket{x}$ of a quantum measurement of the observable $\mathcal{M}$. 
\end{defi}

Typical quantum classifier models proposed in the literature \cite{farhi2018classification, Grant2018, schuld2018circuit, havlivcek2019supervised, schuld2019quantum} fit into this framework. Most choose $\mathcal{M} = \sigma_z \otimes \mathbbm{1} \otimes \hdots \otimes \mathbbm{1}$, so that the measurement queries the computational basis state of the first qubit, and train a variational ansatz after the embedding by optimizing a circuit $U(\theta)$. This can be interpreted as selecting the optimal measurement basis by implementing an adjustable measurement $U(\theta)^{\dagger} \mathcal{M} U(\theta)$. However, $U$ is a linear transformation on the embedded inputs $\ket{x}$, which means that most of the nonlinear discriminating power is determined by the embedding, not the measurement. Note that the quadratic form $\bra{x} \mathcal{M}\ket{x}$ of a measurement introduces a square nonlinearity in $\ket{x}$.\\

To draw connections to quantum state discrimination, one can also define the classification problem in the Hilbert space of the embedding, a version we will call \textit{quantum binary classification}:

\begin{defi} \label{Def:qclass}
For a given embedding $\Phi$, \textbf{quantum binary classification} is the problem of assigning $\ket{x}$ to one of two ``data-ensembles'' of quantum feature states, $\rho = \frac{1}{M_a}\sum_{a} \ketbra{a}{a} $ or $\sigma = \frac{1}{M_b}\sum_{b} \ketbra{b}{b} $. 
\end{defi}

Here, $\rho$ and $\sigma$ are mixed states that describe the process of selecting an embedded data point $\ket{a}, \ket{b}$ with uniform probability from a training set.\\

In the main paper, we discuss two different quantum classifiers (which have also been discussed in other contexts, see \cite{gambs2008quantum, sergioli2017quantum, blank2019quantum, schuld2017implementing}). 

\begin{defi}\label{Def:hel}
Let $\rho - \sigma = \sum_j \lambda_j \ketbra{d_j}{d_j}$ be the difference of two data ensembles $\rho, \sigma$ expressed in the diagonal basis $\{\ket{d_j}\}$ with eigenvalues $\lambda_j$, and let
$$\Pi_+ = \sum\limits_{\lambda_j > 0} \ketbra{d_j}{d_j}, \quad \Pi_- = \sum\limits_{\lambda_j < 0} \ketbra{d_j}{d_j}$$
be projection operators, where the sum runs over all $j$ for which the eigenvalue $\lambda_j$ is positive or negative, respectively. We define the \textbf{Helstr\o{}m classifier} as the quantum classifier $f_{\rm hel}(x) = \bra{x} \Pi_+ - \Pi_- \ket{x}$.
\end{defi}

\begin{defi}\label{Def:fid}
For two data ensembles $\rho, \sigma$, the \textbf{fidelity classifier} is defined as $ f_{\rm fid}(x) = \bra{x} \rho - \sigma \ket{x}$. It measures the fidelity or overlap between data states, since $ \bra{x} \rho - \sigma \ket{x}=  \frac{1}{M_a}\sum_{a} | \braket{a}{x}|^2 - \frac{1}{M_b}\sum_{b} | \braket{b}{x}|^2$.
\end{defi}

\section{Quantum classifiers and state discrimination}\label{App:state_dis}

The problem of quantum classification in Definition \ref{Def:qclass} has been extensively studied in the context of \textit{quantum state discrimination} (i.e., \cite{bae2015quantum}). Two techniques are popularly used in this regard; minimum-error state discrimination and unambiguous state discrimination. In the first case, an assignment of $\ket{x}$ to either $\rho$ or $\sigma$ has a non-zero probability of being erroneous. In the second case, the assignment is promised to be either correct or inconclusive  \cite{chefles2000quantum}.\\

While the problem underlying quantum classification and state discrimination is essentially the same, the two frameworks pose very different requirements to what constitutes a desirable solution. For example, most of the literature in state discrimination asks for the discrimination power of a single measurement. To identify an unknown state from two sets of states, the optimal measurement for single-shot state discrimination is known to be a Helstr\o{}m measurement \cite{helstrom1969quantum}, and the smallest probability of making a wrong guess is governed by the Holevo-Helstr\o{}m theorem. \\

In contrast, a quantum classifier can prepare and measure quantum states multiple times, which requires to run a fixed embedding circuit on a quantum computer, an operation that is relatively cheap to repeat on typical quantum computing platforms. With multiple runs, we can estimate the deterministic expectation $\bra{x} \mathcal{M}\ket{x}$ of the measurement observable $\mathcal{M}$, which renders single-shot or minimum-error considerations meaningless. Conversely, the goal of classification is to generalize, which means to minimize the probability of misclassifying new data points. As a proxy, one minimizes a loss on a set of training examples to find a decision boundary between classes. For a fixed embedding, different measurements correspond to different decision boundaries. As a result, the notion of a ``best'' or ``optimal'' measurement for classification depends on the chosen embedding and loss.\\

In the remainder of the Appendix we present quantum circuits for the \textit{fidelity} and \textit{Helstr\o{}m} classifiers. While both can be \textit{efficiently} implemented (i.e., the number of gates does not grow with the dimension of the Hilbert space in which the data is embedded), we motivate in Appendix \ref{App:hel} how an implementation of the Helstr\o{}m circuit with optimal sample complexity still requires a large number of copies of registers containing the training data, and is therefore not feasible for near-term quantum computers.

\section{The fidelity classifier}\label{App:fid}

Since 
$$\bra{x} \rho - \sigma \ket{x} = \frac{1}{M_a} \sum_a |\braket{x}{a}|^2 - \frac{1}{M_b} \sum_b |\braket{x}{b}|^2,$$
the fidelity classifier can be constructed from evaluating the overlaps $|\braket{x}{a}|^2, |\braket{x}{b}|^2$ -- an operation that requires very few resources on a quantum computer. \\

\begin{figure}
\centering
\includegraphics[width=0.3\textwidth]{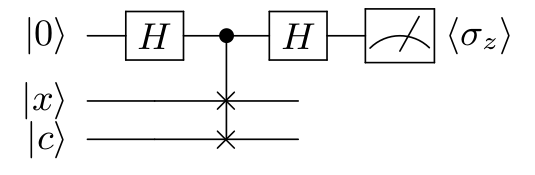}
~~~~~~~~~~~~~~~~~~~ 
\includegraphics[width=0.38\textwidth]{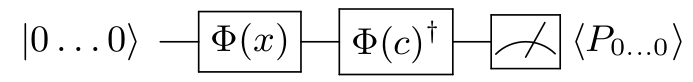}

\caption{Two different routines to measure the overlap of two quantum states. Left: A SWAP test. The third register contains embedded inputs $c = \{a, b\}$ sampled from either class $A$ or $B$. Right: If one can implement an inverse embedding $\Phi^{\dagger}$, the ``inversion test'' measures the overlap $\Phi(c)^{\dagger}\Phi(x)\ket{0 \ldots 0}$ with $\ket{0 \ldots 0}$, using a projector $P_{0 \ldots 0}$ onto the $\ket{0 \ldots 0}$ subspace (and $c=a$ or $c=b$) \cite{havlivcek2019supervised}.  }
\label{Fig:circuit_fid}
\end{figure}

Figure \ref{Fig:circuit_fid} shows two different routines to measure overlaps of two quantum states $\ket{x}$ and $\ket{c}$ represented by $n$ qubits, where $\ket{c} = \{\ket{a}, \ket{b}\}$ in this application. The left circuit implements a SWAP test, which evaluates overlaps of two quantum states using $2n+1$ qubits and has no general requirements for the feature map $\Phi$. Measuring the ancilla multiple times allows us to estimate the expectation of the first qubit, $\langle \sigma_z\rangle = |\braket{c}{x}|^2$. As mentioned in Ref \cite{havlivcek2019supervised}, if one can implement the inverse  $\Phi(x, \theta)^{\dagger}$ of the embedding circuit $\Phi(x, \theta)$, the right ``inversion test'' circuit can be used to significantly save resources.\\
 
Using $\rho - \sigma = \sum_j \lambda_j \ketbra{d_j}{d_j}$,  the expression
$$ \bra{x} \rho - \sigma \ket{x} = \sum_{j} \lambda_j |\braket{d_{j}}{x}|^2 $$
reveals that the fidelity classifier is equivalent to the Helstr\o{}m classifier (compare Eq. (\ref{Eq:exp_hel})), but instead of only considering the sign of the eigenvalues $\lambda_j$, it also considers their magnitude. 

\section{The Helstr\o{}m classifier}\label{App:hel}

An implementation for a Helstr\o{}m measurement has been suggested in the context of quantum principal component analysis (QPCA) \cite{lloyd2014quantum}, a routine that ``simulates'' a density matrix to extract its eigenvalues. The scheme requires roughly $\mathcal{O}(R n)$ gates, where $R$ is the rank of the exponentiated matrix, and $n$ is the number of qubits used to represent the eigenstates. QPCA was shown to be optimal with respect to the required number of copies of the simulated density matrix \cite{kimmel2017hamiltonian}. Ref \cite{kimmel2017hamiltonian} also shows that the framework easily extends to simulating the difference of two density matrices. This can be used to exponentiate $m = \rho - \sigma$, and extract the sign of $m$'s eigenvalues $\lambda_j$ with optimal sample complexity.  \\

Whether Helstr\o{}m measurements can be efficiently implemented was an open question in \cite{gambs2008quantum}. Using the QPCA routine, an efficient implementation would require the rank $R$ to grow at most polynomially with the number of qubits $n$. We show here that this is indeed the case for classification in our framework. As explained in more detail below, in every run of the circuit one constructs the measurement using two pure states $\ketbra{a}{a}, \ketbra{b}{b}$ sampled from the data ensemble, and $\ketbra{a}{a} - \ketbra{b}{b}$ has a maximum rank of $2$. However, the QPCA implementation of the Helstr\o{}m classifier is unlikely to be used on near-term quantum computers, since typical implementations of the phase estimation subroutine all require a large number of registers coherently prepared in states $\ket{a}, \ket{b}$.

\subsection{Concept}

The Helstr\o{}m classifier can be implemented by a quantum circuit that results in the following expectation of a single-qubit Pauli-Z observable:
\begin{equation} 
\langle \sigma_z \rangle = \sum_j \mathrm{sign}(\lambda_j) |\braket{d_j}{x}|^2.  
\label{Eq:exp_hel}
\end{equation}

Since
$$ \sum_j \mathrm{sgn}(\lambda_j) |\braket{d_j}{x}|^2 = \sum\limits_{\lambda_j > 0}  |\braket{d_j}{x}|^2 - \sum\limits_{\lambda_j < 0}  |\braket{d_j}{x}|^2 = \mathrm{tr}\{ \Pi_+ \ketbra{x}{x}\} - \mathrm{tr}\{ \Pi_- \ketbra{x}{x}\},$$
the expectation $\langle \sigma_z \rangle$ carries all necessary information for the assignment: if the expectation is positive -- or equivalently, the corresponding qubit is more likely to be measured in state $\ket{0}$ than $\ket{1}$ -- we assign the new input to $\rho$, and otherwise to $\sigma$. The challenge of implementing this result in a quantum circuit is to evaluate the sign function, which is non-smooth.\\

The QPCA routine suggests a method to implement the Helstr\o{}m measurement using a combination of density-matrix exponentiation (DME) and quantum phase estimation (QPE). Roughly speaking, DME prepares a state 
$$e^{- i \delta (\rho-\sigma)} \ket{x}\ket{0\ldots0} = \sum_j \braket{d_j}{x}e^{- i \delta \lambda_j} \ket{d_j}\ket{0\ldots0} $$ 
up to error $\mathcal{O}(\delta^2)$. DME can be simulated for longer times $ k \delta$ and up to error $\textit{O}(k \delta^2)$ if $k$ data copies of $\rho, \sigma$ are used. By choosing specific values in $[0, 2\pi]$ for $k \delta$, introducing a phase shift (to resolve the sign of the phase) and controlling the operation on $L$ ancilla qubits, one can use DME as a subroutine of standard QPE \cite{nielsen2002quantum}. QPE creates a superposition of phases encoded in the computational basis states of the ancillas. The first qubit in the $j$th basis state is zero if the $j$th eigenvalue is negative, and one otherwise. The expectation of this qubit is consequently proportional to the desired value for $\langle \sigma_z \rangle$. \\

Measuring the first qubit in the QPE register hence resolves the sign of the eigenvalues of $\rho-\sigma$, weighted by the overlap $|\braket{d_j}{x}|^2$, as specified in Eq. (\ref{Eq:exp_hel}). Note that one cannot make use of slimmer iterative QPE schemes (see \cite{kimmel2017hamiltonian}), since the result of the computation is not a single phase, but the expectation of a qubit which depends on a ``superposition of extracted phases''.

\subsection{Sampling pure training states}\label{App:sampling}

The observable $\Pi_+ - \Pi_-$ of a Helstr\o{}m measurement is formally constructed from  $\rho - \sigma$. However, when implementing the QPCA-based measurement circuit, each run uses samples $a, b$ from the two classes, so that in every run the measurement is based on the pure states $\ketbra{a}{a} - \ketbra{b}{b}$. One then averages the expectation values estimated for every $\ket{a}, \ket{b}$ pair, $\langle \sigma_z \rangle_{a,b}$, to get the overall expectation from Eq. (\ref{Eq:exp_hel}),
$$ \langle \sigma_z \rangle = \frac{1}{M_a} \frac{1}{M_b} \sum_a \sum_b \langle \sigma_z \rangle_{a,b}. $$

To see that averaging over pure-state expectations is formally equivalent to the mixed-state expectation, consider a circuit $W$ applied to an initial state $\ket{\psi} = \ket{...} \otimes \ket{a} \otimes \ket{b}$. Using density-matrix notation $\eta = \ketbra{\psi}{\psi}$, the final expectation value of an observable $\mathcal{M}$ is given by $ \text{tr} \{ W \eta W^{\dagger} \mathcal{M}  \} $. Averaging over several runs of the circuit with $a, b$ sampled with probabilities $p_a, p_b$, we get the ``classical expectation of the quantum expectation''
$$ \sum_a \sum_b p_a p_b \; \text{tr} \{ W \eta W^{\dagger} \mathcal{M}  \}  = \text{tr} \{ W \left( \sum_a \sum_b p_a p_b \eta \right) W^{\dagger} \mathcal{M}  \},$$
where
$$
\sum_a \sum_b p_a p_b \eta  =  \ketbra{...}{...} \otimes \sum_a p_a \ketbra{a}{a} \otimes \sum_b p_b\ketbra{b}{b} = \ketbra{...}{...} \otimes\rho \otimes \sigma .
$$

Note that reconstructing the ensembles by averaging over pure states means that we cannot use approximate schemes that scale the result of a single run by an arbitrary value. For example, a single-qubit pointer method evaluates $ \mu \langle \sigma_z \rangle_{a,b}$ up to an unknown factor $\mu$, which skews the overall average. This is the reason why many ``shortcuts'' for quantum phase estimation cannot be used to reduce the resources of the QPCA-based Helstrom measurement circuit.

\subsection{The Helstr\o{}m measurement is efficient }

The matrix $m = \ketbra{a}{a} - \ketbra{b}{b}$ has properties that allow for a particularly efficient implementation of the Helstr\o{}m measurement with the QPCA strategy. If $\ket{a} \neq \ket{b}$, $m$ is of rank $2$. Furthermore, since $\tr(\ketbra{a}{a}- \ketbra{b}{b}) = \tr(\ketbra{a}{a}) - \tr(\ketbra{b}{b}) =  0$, $m$ has trace zero. The only two non-zero eigenvalues are therefore $\pm \lambda$. Their magnitude depends on the overlap $\braket{a}{b}$; for $\braket{a}{b} = 0$, $\lambda=1$, and for $\braket{a}{b} = 1$ we have $\lambda=0$. \\

As a result, the rank of the matrix we exponentiate is constant and small, while the average size of the eigenvalue depends on how well the embedding is trained. A successful embedding decreases the inter-class overlaps and therefore increases the absolute value of the eigenvalue on average, making it easier to estimate its sign via quantum phase estimation.\\

However, and reflecting the huge gap between traditional quantum computing and the requirements of near-term hardware, even if the Helstr\o{}m measurement can be efficiently implemented, the required number of registers prepared in $\ket{a}, \ket{b}$ to simulate $\ketbra{a}{a} - \ketbra{b}{b}$ for times in $[0, 2\pi]$ as a subroutine to quantum phase estimation is still prohibitive for near-term quantum computing. For example, for time $1$, $\delta = 0.01$ and $L=10$ QPE qubits, we need $k=100$ copies of $\ket{a}, \ket{b}$, which translates into at least $2000n$ qubits for embeddings that use $n$ qubits.

\end{document}